\documentclass[conference]{IEEEtran}
\IEEEoverridecommandlockouts

\usepackage{cite}
\usepackage{amsmath,amssymb,amsfonts}
\usepackage{algorithmic}
\usepackage{graphicx, caption}
\usepackage{threeparttable}
\usepackage{subfigure}
\usepackage{textcomp}
\usepackage{xcolor}
\usepackage{etoolbox}
\def\BibTeX{{\rm B\kern-.05em{\sc i\kern-.025em b}\kern-.08em
    T\kern-.1667em\lower.7ex\hbox{E}\kern-.125emX}}

\title{Accuracy and Application Scope Analysis for Linearized Branch Flow Model in Radial Distribution Systems}

\author{\IEEEauthorblockN{Hanyang Lin, Xinwei Shen*}
\IEEEauthorblockA{Tsinghua-Berkeley Shenzhen Institute\\Shenzhen International Graduate School\\
Tsinghua University\\
Shenzhen, China}
\and
\IEEEauthorblockN{Qinglai Guo, Hongbin Sun}
\IEEEauthorblockA{Department of Electrical Engineering\\
Tsinghua University\\
Beijing, China}
\thanks{Hanyang Lin is also with Electrical and Electronic Engineering Department, Imperial College London at London, United Kingdom. Corresponding author: Xinwei Shen, e-mail: sxw.tbsi@sz.tsinghua.edu.cn. This work is supported by the National Natural Science Foundation of China (No. 52007123).}}
    
\makeatletter
\patchcmd{\@maketitle}
  {\addvspace{0.5\baselineskip}\egroup}
  {\addvspace{-1.6\baselineskip}\egroup}
  {}
  {}
\makeatother

\begin{document}

\IEEEtitleabstractindextext{
\noindent\begin{abstract}
An in-depth analysis of linearized branch flow (LBF) model considering current injection and absolute value of impedance is proposed in this paper. The form of LBF model is based on two equations: the current injection to meet KCL and the voltage drop to meet KVL. By representing the absolute value of complex load power with the current injection, LBF model is much simpler than alternating current power flow (ACPF) model. The results on theoretical analysis and numerical studies show that LBF exhibits the high accuracy in bus voltage magnitude but a poor performance in branch flow. Moreover, LBF is also compared with fast decoupled linearized power flow (FDLPF) model to verify its efficiency, thus proving its superiority for fast evaluation of large-scale distribution systems with high accuracy in voltage magnitude. Finally, this paper analyzes three factors to lower LBF's errors of branch flow, as well as LBF's possible application scope.  
\end{abstract}\vspace{5pt}

\noindent\begin{IEEEkeywords}
 radial distribution system, power flow analysis, linear power flow model, voltage magnitude, branch flow
\end{IEEEkeywords}}
\maketitle
\IEEEdisplaynontitleabstractindextext
\IEEEpeerreviewmaketitle

\section{Introduction}
Power flow model is an essential component of planning and optimization of increasingly complex and uncertain distribution networks, including network reconfiguration and economic dispatch. As the traditional alternating current power flow (ACPF) is nonlinear, inefficient and usually makes the optimization problem non-convex, it is important to introduce a linear power flow (LPF) model with high efficiency and acceptable accuracy for distribution networks. Direct current power flow (DCPF) is a widely used LPF model for transmission networks, but it is not directly applicable to distribution networks\cite{purchala2005usefulness}. DCPF cannot be used in some applications like automatic voltage control\cite{sun2012adaptive}. In this case, different LPF models have been proposed in previous works.

According to the selection of state variables, there exist two models: bus injection model and branch flow model\cite{yuan2016novel}. Most LPF models\cite{yuan2016novel,garces2015linear,gharebaghi2019linear} base on bus injection model which focuses mainly on node variables. As for branch flow model\cite{haffner2008multistage,yang2016state,teng2003direct}, it focuses primarily on power flows of branches. Besides, there are warm-start and cold-start models associated with branch flow models. Warm-start models focus the operating points with the initial operating points in order to linearize ACPF model\cite{chen2017robust}. Cold-start models do not require the initial operating points, which means that the accuracy of cold-start models is worse than warm-start models\cite{9281770}. In addition, different start models have different application situations. For example, a well-known linear branch flow model called Simplified DistFlow model\cite{baran1989network} is a cold-start model with a poor accuracy. As for warm-start models, the three-phase linear branch flow model in \cite{wang2017linear} formulated the model of on-load-tap-charger in transformers with a high accuracy of branch flows. Also, the decoupled linearized power flow (DLPF) in \cite{yang2016state} is state independent and has a high accuracy in voltage magnitude, it also decreased the computing time by proposing the fast decoupled linearized power flow (FDLPF), which is one of the comparisons of LBF model in this paper.

Also, studies of ACOPF with convex relaxation became important. A branch flow model using a second-order cone program (SOCP) to solve efficiently for both mesh and radial networks is proposed in \cite{farivar2013branch}. Reference\cite{gan2014exact} proved that a global optimum of OPF can be obtained by solving SOCP under a mild condition and the relaxation is exact for radial networks. ACOPF based on SOCP to solve a series of power flow sets is used as the another comparison of LBF model in this study.

This paper analyzed a linearized disjunctive model in-depth adapted from alternating current load flow linearized network model given in \cite{haffner2008multistage}, referred to linearized branch flow (LBF). Also, this LBF model is explained in \cite{teng2003direct} as a direct approach for power flow solutions. However, previous studies did not analyze LBF's accuracy of branch flow (neither active power flow nor current), which is the key result of LPF model. Therefore, this paper further analyzed the accuracy of node voltage and branch flow in details, to give a clear description of its performance. Besides, this paper discussed the error's affecting factors and found LBF's application potentials in distribution network expansion planning problems. 
The major contributions of the paper are two-fold:

(i)	LBF is compared with linear and nonlinear power flow models to show that LBF is the simplest power flow model and much more efficient to solve with lowest computing time in large-scale distribution system tests. And the accuracy of LBF is acceptable in some cases, while in some other cases and application scenes are not. 

(ii) LBF's errors and LBF's application scope are discussed based on analytical equations as well as case studies in this study. We found that LBF can provide lower bounds for nodes' voltage and upper bounds for branch flows given a proper scaling coefficient in current injection, thus making it a warm start for some distribution system optimization problems (e.g. distribution system expansion planning, DSEP).

The remaining of this paper is organized as follows: Section \uppercase\expandafter{\romannumeral2} is the formulation of linearized branch flow model. Section \uppercase\expandafter{\romannumeral3} uses the theoretical analysis to analyze the potential errors and discuss factors that would lower errors. In Section \uppercase\expandafter{\romannumeral4}, the proposed LBF, FDLPF and the ACPF based on SOCP are compared in three test systems.

\section{Problem Formulation}

\subsection{Basic Model of ACOPF Based on SOCP}

According to the branch flow model first modified in\cite{farivar2013branch} and showed in Fig 1, $S_{i}$ is the complex power of load at node $i$, $S_{ij}$ is the complex power flow from node $i$ to node $j$, $l_{ij}$ is the magnitude squared of current from node $i$ to node $j$, and $v_{j}$ is the magnitude squared of voltage at node $j$. 
\begin{figure}[ht!]
\captionsetup{singlelinecheck = false, format= hang, justification=raggedright, font=footnotesize, labelsep=space}
\centering
\vspace{-0.2cm}
\includegraphics[scale=0.45]{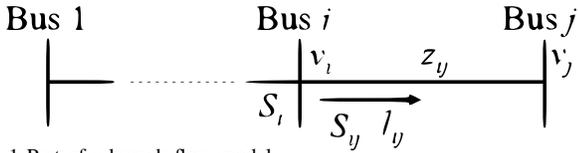}
\setlength{\abovecaptionskip}{-1.5pt}
\setlength{\belowcaptionskip}{-1.5pt}
\caption{Part of a branch flow model}
\label{fig}
\end{figure}

We assume that the complex voltage at node 1, which is the slack bus, is given and the complex net generation $s_{1}$ is a variable. Other power injections $s:=(s_{i},i=2,...,n)$ are given for power flow (PF) analysis, as for optimal power flow (OPF) analysis, $s$ are control variables. Given $z:=(z_{ij},(i,j)\in E,z_{i},i\in N)$, $V_{0}$ and bus power injections $s$ satisfy the Ohm's law:
\begin{equation}
    V_{i} - V_{j} = z_{ij}I_{ij}, 
\end{equation}
the definition of branch flow:
\begin{equation}
    S_{ij}=V_{i}I_{ij}^{*}
\end{equation}
and complex power balance at each bus:
\begin{equation}
    \sum\limits_{k:j\to k}S_{jk}-\sum\limits_{i:i\to j}(S_{ij}-z_{ij}|I_{ij}|^2 )+y_{j}^* V_{j}^2=s_{j}
\end{equation}
where $y_{i}$ is the admittance at bus $j$, and $S_{jk}$ means the complex power plow from node $j$ to other connecting nodes through branches. Equations (1)-(3) are the branch flow model which are nonlinear equations. The solutions of equations (1)-(3) are called branch flow solutions with respect to a given $s$.

Connecting (2) and (3) with real value of variables, and taking the magnitude squared of equation obtained by substituting (2) into (1), therefore we have:
\begin{equation}
    p_{j}= \sum\limits_{k:j\to k}P_{jk}-\sum\limits_{i:i\to j}(P_{ij}-r_{ij}l_{ij} )+g_{j}v_{j}
\end{equation}
\begin{equation}
    q_{j}= \sum\limits_{k:j\to k}Q_{jk}-\sum\limits_{i:i\to j}(Q_{ij}-x_{ij}l_{ij} )+b_{j}v_{j}
\end{equation}
\begin{equation}
    v_{j}=v_{i}-2(r_{ij}P_{ij}+x_{ij}Q_{ij})+(r_{ij}^{2}+x_{ij}^{2})l_{ij}
\end{equation}
\begin{equation}
    l_{ij}=\frac{P_{ij}^2+Q_{ij}^2}{v_{i}}
\end{equation}
where $p_{j}$ and $q_{j}$ are the active power and reactive power at node $j$, $P_{jk}$ and $Q_{jk}$ are the active power and reactive power flow from node $j$ to other connecting nodes through branches, and $P_{ij}$ and $Q_{ij}$ are the active power and reactive power flow from node $i$ to node $j$.

We call equations (4)-(7) the branch flow model, however the model is still non-convex and cannot be solved directly. Under this condition, second order cone constraint (SOCP) is introduced in (7) to make the feasible set convex.
\begin{equation}
    l_{ij}\geq \frac{P_{ij}^2+Q_{ij}^2}{v_{i}}
\end{equation}

Since the objective function of ACOPF problem is loss minimization, like all PF problems, thus this function is convex and this model is a conic optimization. Then the relaxation is exact in radial distribution networks for both PF and OPF problems\cite{gan2014exact}. With the given $s$, we actually use ACOPF based on SOCP model to solve a fixed ACPF problem. In the case study section, we extend this model to a long-term horizon with $t$ dimensions.

\subsection{Linearized Branch Flow model}
The proposed LBF model are used for the problem of distribution systems with a long-term horizon, and some factors need to be taken into account:
\begin{itemize}
\item The problem has a long-term horizon which is $T$ stages and all the relevant variables are linked to each stage. Each stage has different capacity of node loads and branch impedance.
\item Each node has two continuous variables: the former is the magnitude of node voltage and the latter is the current injection; each branch has one continuous variable, current flow $f$;
\item A coefficient a, 1.08, is multiplied with load $d_{j,t}$ to scale the current injection and observe the variation of branch flows and node voltages.
\end{itemize}
\begin{figure}[ht!]
\captionsetup{singlelinecheck = false, format= hang, justification=raggedright, font=footnotesize, labelsep=space}
\centering
\vspace{-0.2cm}
\setlength{\abovecaptionskip}{-1.5pt}
\setlength{\belowcaptionskip}{-1.5pt}
\includegraphics[scale=0.43]{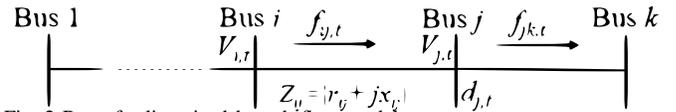}
\caption{Part of a linearized branch flow model}
\label{fig}
\end{figure}

Consider a part of radial network structure, as shown in Fig.2, then the following expressions (in p.u.) are obtained from Kirchhoff’s current law (KCL) and Kirchhoff’s voltage law (KVL):
\begin{equation}\label{equ:1}
    \begin{split}
         f_{ij,t} &= a*d_{j,t} + f_{jk,t} \\
\end{split}
\end{equation}
\begin{equation}\label{equ:1}
    \begin{split}
    \Delta V_{ij,t} &= V_{i,t} - V_{j,t} = Z_{ij}f_{ij,t}
\end{split}
\end{equation}
where $f_{ij,t}$ indicates current of branch between node $i$ and node $j$ in the fixed network at stage $t$, $f_{jk,t}$ is the current from node $j$ to other connecting nodes, taking node $k$ as an example and $d_{j,t}$ stands for apparent power of load in node $j$ at stage $t$. $a$ is the multiple coefficient and is set to 1 for original LBF model. $\Delta V_{ij,t}$ means the difference of the magnitude of voltage between node $i$ and node $j$ at stage $t$, $Z_{ij}$ is the absolute value of impedance of Branch from node $i$ and node $j$. 
As for the apparent power of load:
\begin{equation}
    d_{i,t} = |P_{di,t}+jQ_{di,t}|
\end{equation}
where $P_{di,t}$ and $Q_{di,t}$ represent the active power and reactive power of load in node $i$ at stage $t$. In (9), the absolute value of complex power of load is regarded as the outflow current, so the current injections satisfy KCL; in (10), the voltage drop is the product of the absolute value of branch impedance and current, which satisfies KVL. 

This paper also takes fast decoupled linearized power flow model (FDLPF) as the comparison of LBF model\cite{yang2016state}. FDLPF model is designed by developing a fast approximation of the matrices used in DLPF model to have a lower computing time.

\section{Theoretical Analysis for LBF's Errors}
There are two main obvious errors, error of voltage magnitude and error of branch flow, are associated with LBF model. This section will discuss how these errors occur and analyze the factors that would affect errors. 

\subsection{Error of Node Voltage}\label{AA}
The error of node voltage is discussed in a two-bus system, including a substation node 0 and a load node 1, as shown in Fig.3. $\Delta V_{LBF}$ and $\Delta V_{AC}$ are the voltage drop between two nodes in LBF and ACPF model, respectively. As LBF is a lossless model and the voltage drop caused by the power loss in ACPF model can be ignored in this two-bus system, the power flow is the load power $\dot{S}_{1}$ at bus 1 in both two models. The branch current $\dot{I}_{12}$ in ACPF is calculated as belows:
\begin{figure}[ht!]
\captionsetup{singlelinecheck = false, format= hang, justification=raggedright, font=footnotesize, labelsep=space}
\centering
\vspace{-0.41cm}
\setlength{\abovecaptionskip}{-1.5pt}
\setlength{\belowcaptionskip}{-1.5pt}
\includegraphics[scale=0.42]{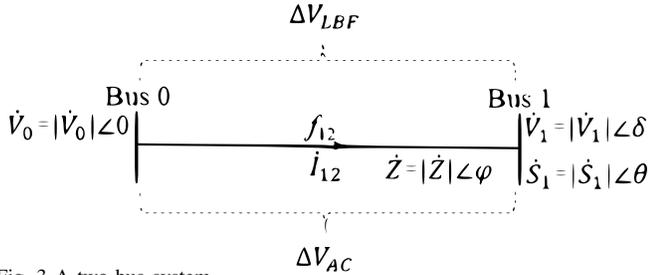}
\caption{A two-bus system}
\label{fig}
\end{figure}\vspace{-0.4cm}
\begin{equation}
    \dot{I}_{12}=(\dot{S}_{1}/\dot{V}_{1})^{*}=\bigl(|\dot{S}_{1}|/|\dot{V}_{1}|\angle(\theta-\delta)\bigr)^{*}
\end{equation}
where $|\dot{V}_{1}|$ is around 1 p.u, thus we assume that $|\dot{V}_{1}|$ is 1 p.u and calculate $\dot{V}_{1}$:
\begin{equation}
\begin{split}
     \dot{V}_{1}&=\dot{V}_{0}-\dot{I}_{12}\dot{Z}\\&=\bigl(\dot{V}_{0}-|\dot{S}_{1}||\dot{Z}|cos(\gamma)\bigr)-j|\dot{S}_{1}||\dot{Z}|sin(\gamma)\\&=(\dot{V}_{0}-a)-jb   
\end{split}
\end{equation}
where $\gamma$ is $\varphi-\theta+\delta$, $a$ denotes $|\dot{S}_{1}||\dot{Z}|cos(\gamma)$ and $b$ represents $|\dot{S}_{1}||\dot{Z}|sin(\gamma)$. And we can calculate $\Delta V_{AC}$ and $\Delta V_{LBF}$:
\begin{equation}
    \Delta V_{AC}=|\dot{V}_{0}|-|\dot{V}_{1}|=|\dot{V}_{0}|-\sqrt{(\dot{V}_{0}-a)^{2}+b^2}
\end{equation}
\begin{equation}
    \Delta V_{LBF}=f_{12}|\dot{Z}|=|\dot{S}_{1}||\dot{Z}|
\end{equation}

In order to compare the value of these two voltage drop, we assume $\Delta V_{LBF}$ is larger than $\Delta V_{AC}$ first:\vspace{-2pt}
\begin{equation}\nonumber
\Delta V_{LBF}-\Delta V_{AC}=|\dot{S}_{1}||\dot{Z}|-\bigl(|\dot{V}_{0}|-\sqrt{(\dot{V}_{0}-a)^{2}+b^2}\bigr)\geq{0}
\end{equation}
then we have:\vspace{-2pt}
\begin{equation}\nonumber
-(|\dot{S}_{1}||\dot{Z}|-|\dot{V}_{0}|)\leq \sqrt{(\dot{V}_{0}-a)^{2}+b^2}
\end{equation}
note that $|\dot{V}_{0}|$ is always bigger than $|\dot{S}_{1}||\dot{Z}|$, thus take the square of both sides of this equation:\vspace{-2pt}
\begin{equation}\nonumber
 (|\dot{S}_{1}||\dot{Z}|-|\dot{V}_{0}|)^{2}\leq (\dot{V}_{0}-a)^{2}+b^2
\end{equation}
and the result of the equation above is:\vspace{-2pt}
\begin{equation}\nonumber
 |\dot{S}_{1}||\dot{Z}|\geq|\dot{S}_{1}||\dot{Z}|cos(\gamma)
\end{equation}
and hence:\vspace{-2pt}
\begin{equation}\nonumber
 1 \geq cos(\gamma)
\end{equation}
this condition, $1 \geq cos(\gamma)$, always holds and all the processes of this proof are reversible, thus the assumption, $\Delta V_{LBF} \geq \Delta V_{AC}$, holds. Therefore we prove that $\Delta V_{LBF}$ is larger than $\Delta V_{AC}$, which means the node voltages calculated by LBF are smaller than the results by ACPF given the same voltage of substation node and mild assumption that $|\dot{V}_{1}|\approx 1$. This is where the error of node voltage originates from.
\subsection{Error of Branch Flow}
As described in the Model Formulation, the branch current, $I$, calculated from ACPF based on SOCP is associated with voltage drop and resistance of each branch. LBF uses the absolute value of complex power to represent branch currents, $f$, which is calculated by the summation of rest nodes' active power and reactive power, including losses. The error of current would be affected significantly by reactive load power in certain nodes and the resistance of each branch. Thus, three factors obtained from error analysis are considered in this study to find the application scope of LBF:
\begin{itemize}
\item \emph{Different $P/Q$ ratios:} $P/Q$ ratio means the value of $P/Q$ at nodes or buses. According to equations (9)-(11), the different $P/Q$ ratios can affect the value of $f_{ij,t}$, thus affecting the voltage drop. If the value of $P/Q$ ratio is low, which means that the value of reactive power may be equal or larger than active power of load at certain nodes, the difference between $f$ and $I$ is large, and vice versa. So the errors of branch flow and node voltage would be seriously affected by $P/Q$ ratios in load nodes.
\item \emph{Different $r/x$ ratios:} $r/x$ ratio means the value of $r/x$ in the branch connecting nodes. As the different values of resistance would affect $I$, the error of branch flow would be small if $r/x$ ratio of the branch situation is large. Besides, according to equation (10), different $r/x$ ratios can affect the voltage drop, so both the error of node voltage and branch flow would be affected by voltage drop. It should be noted that, in a mid-level distribution system (e.g. 10 kV in China and 12.6 kV in U.S.), $r/x$ ratios are relatively fixed within a certain range.
\item \emph{Different load levels:} As the load level of the whole system increases, both the errors of node voltage and branch flow will usually increase. However, if the load levels at some nodes are stressed, the error of branch flow will decrease since the stressed value of $P$ and $Q$ alleviate the effect of $r$.
\end{itemize}
\section{Case Studies}
The case studies are mainly performed on the 33-bus \cite{baran1989network}, 69-bus and 141-bus radial distribution systems. The base voltage and initial load conditions can be checked in MATPOWER. In all tests, the voltage magnitude of the substation node is set as 1.05 p.u and all the units are set as p.u.
\subsection{Simulation Results}\label{AA}
Table \uppercase\expandafter{\romannumeral1} compares the performances and characteristics of different power flow models on three test systems. In the table, we use ACPF short for ACPF based on SOCP, whose results are the benchmark. The error of node voltage and error of branch flow are the average error of three test systems. Also, $n$ means the bus (node) number and $m$ is the branch number.

\begin{table}[ht]
   \centering
 \vspace{-0.2cm}
   \setlength{\abovecaptionskip}{-1.0pt}
   \setlength{\belowcaptionskip}{-1.5pt}
   \captionsetup{labelformat=default,labelsep=space}
   \caption{\\Characteristics of Different Power Flow Models} 
   \label{Non-base}
   \begin{tabular}{cccc}
    \hline
    \hline
     Characteristics & ACPF & FDLPF & LBF \\
    \hline
    Number of Variables & $3n+2m+2$ & $2n+m+2$ & $n+m+1$ \\
    Number of Equations & $3n+3m+1$ & $2n+2m+1$  & $n+2m$ \\
    Error of Node Voltage & $0\%$ & $1.5\%$ & $0.43\%$ \\
    Error of Branch Flow & $0\%$ & $2.74\%$ & $5.16\%$ \\
    \hline
    \hline
   \end{tabular}
  \end{table}
\vspace{-0.3cm} 
According to Table \uppercase\expandafter{\romannumeral1}, LBF has the smallest number of variables and equations, also its equations does not include iteration parts, which means that LBF is the simplest compared to others. FDLPF simplifies the number of variables and equations of ACPF, however it still has a certain amount of complex equations to guarantee its accuracy. As for error of node voltage, FDLPF has a high accuracy in the 33-bus systems, however when the system becomes larger, the error of FDLPF becomes larger than that of LBF. Since the approximation of current by apparent power, LBF has a relatively larger error of branch flow. 

\begin{table}[ht]
   \centering
 \vspace{-0.2cm}
   \setlength{\abovecaptionskip}{-1.0pt}
   \setlength{\belowcaptionskip}{-1.5pt}
   \captionsetup{labelformat=default,labelsep=space}
   \caption{\\ Computing time of Different Power Flow Models} 
   \label{Non-base}
   \begin{tabular}{cccc}
    \hline
    \hline
     Simulation Cases & ACPF(s) & FDLPF(s) & LBF(s) \\
    \hline
    Case33 & $24.731$ & $5.752$ & $1.337$ \\
    Case69 & $54.465$ & $8.371$  & $2.936$ \\
    Case141 & $78.168$ & $16.849$ & $5.699$ \\
    \hline
    \hline
   \end{tabular}
  \end{table}
\vspace{-0.3cm} 

The results of computing times for the large-scale systems of three power flow models are presented in Table \uppercase\expandafter{\romannumeral2}. In this comparison, all three distribution systems are expanded to 100 dimensions, which have different load conditions of nodes and impedance conditions of branches. FDLPF has an improved computing efficiency and preforms a little worse than DC power model\cite{yang2016state}. It is evident that LBF is the most efficient model in large-scale systems. 

\subsection{Error Analysis}\label{AA}
Fig. 4-6 visualizes the errors of node voltage and branch flow of LBF with $'a'$ in eq. (9) as scaling coefficient (called MLBF) and FDLPF in three test systems. In Fig. 4 and 5, MLBF has a relatively larger error of node voltage magnitude compared to FDLPF. However, as shown in Fig. 6, it is obvious that when the system becomes larger, FDLPF loses its accuracy according to the flaws in FDLPF model itself (the linear voltage drop equation). As refer to error of branch flow, MLBF has a relatively larger error compared with FDLPF, and the error of branch flow would increase rapidly if the reactive load power at some nodes is large, theoretically explained in Section \uppercase\expandafter{\romannumeral3}. Note that MLBF exhibits the higher efficiency and accuracy in node voltage magnitudes but less accuracy in branch flow. However, with the help of scaling coefficient $a$ at a proper value (e.g. a =1.08 in this case), MLBF also showed an important feature: the errors of node voltages are always negative, while the errors of branch flows are always positive. Hence MLBF, with a proper scaling coefficient $'a'$ in eq. (9), could produce lower bounds for node voltages and upper bounds for branch flows at a low computation cost (linear programming, with less variables and equations), which are useful for some distribution system optimization problems, e.g. distribution system expansion planning and reconfiguration.

\begin{figure}[ht!]
\captionsetup{singlelinecheck = false, format= hang, justification=raggedright, font=footnotesize, labelsep=space}
\centering
\vspace{-0.46cm}
\setlength{\abovecaptionskip}{-1.5pt}
\setlength{\belowcaptionskip}{-1.5pt}
\includegraphics[width=3.4in]{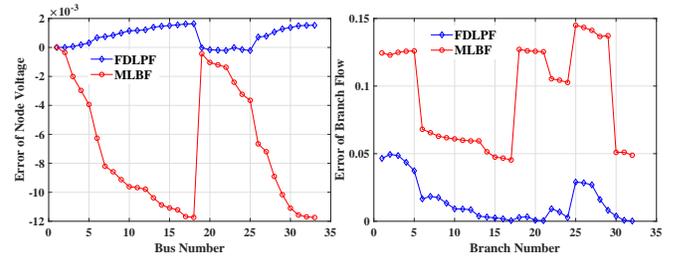}
\caption{Error Comparison between LBF and FDLPF in 33-Bus system}
\label{fig}
\end{figure}\vspace{-6pt}

\begin{figure}[ht!]
\captionsetup{singlelinecheck = false, format= hang, justification=raggedright, font=footnotesize, labelsep=space}
\centering
\vspace{-0.5cm}
\setlength{\abovecaptionskip}{-1.5pt}
\setlength{\belowcaptionskip}{-1.5pt}
\includegraphics[width=3.4in]{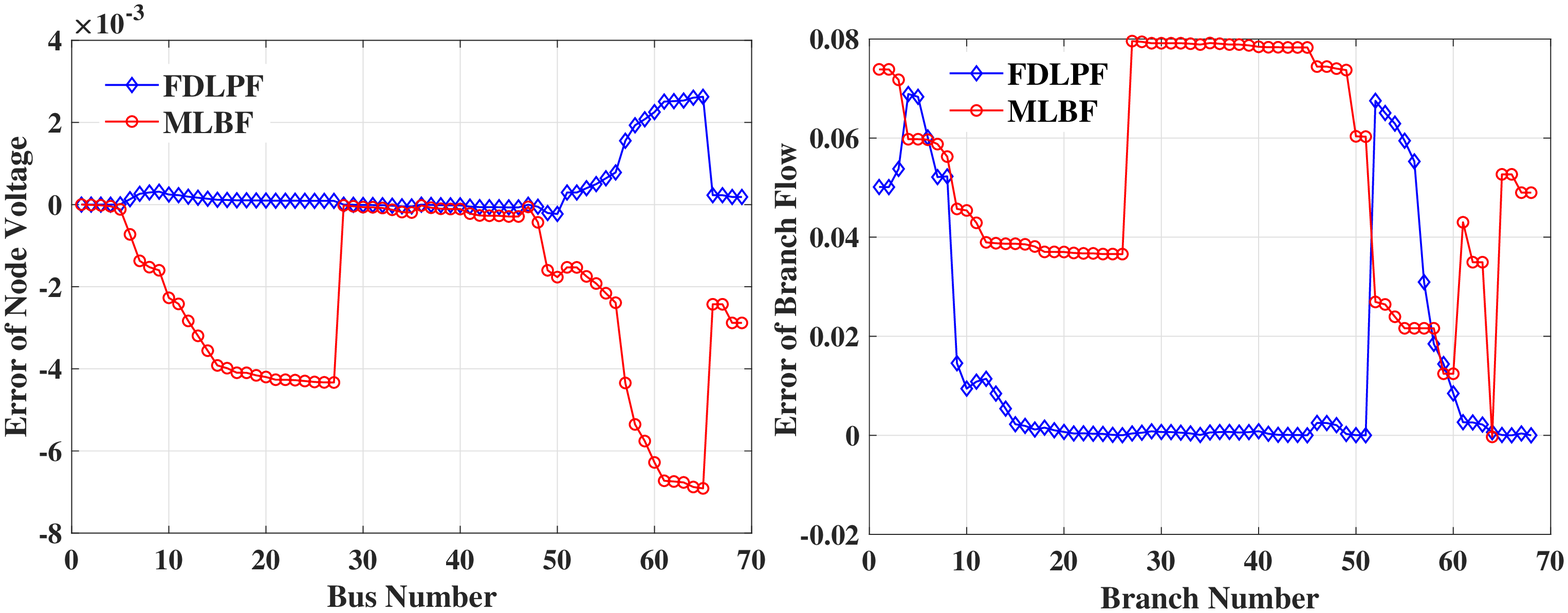}
\caption{Error Comparison between LBF and FDLPF in 69-Bus system}
\label{fig}
\end{figure}
\vspace{-7pt}
\begin{figure}[ht!]
\captionsetup{singlelinecheck = false, format= hang, justification=raggedright, font=footnotesize, labelsep=space}
\centering
\vspace{-0.5cm}
\setlength{\abovecaptionskip}{-1.5pt}
\setlength{\belowcaptionskip}{-1.5pt}
\includegraphics[width=3.35in]{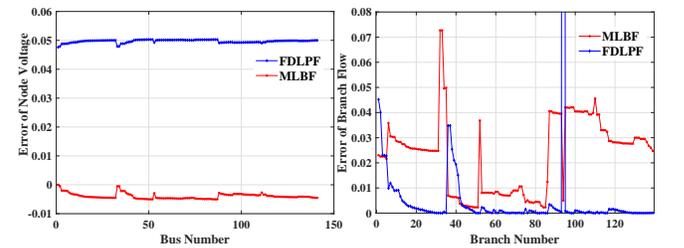}
\caption{Error Comparison between LBF and FDLPF in 141-Bus system}
\label{fig}
\end{figure}\vspace{-8pt}

\subsection{Applicable Scope}\label{AA}
As LBF has the poor performance in branch flow, it is of highly desirable to find the application scope of LBF ($a$ is set to 1). In Fig. 7, if $f$ is larger than $I$, the error would be lower by changing $P/Q$ ratio. Take 33-bus system as an example, error increases significantly from node 25 to node 29, because of a Static Var Compensator at node 30. Different $P/Q$ ratios of load condition at node 30 are set to observe the changes of error. As shown in Fig.7, when $P/Q$ ratio of load condition increases, the error declines. If $f$ is smaller than $I$, the error would be lower by changing $r/x$ ratio. The errors of 69-bus system are negative because the branch resistance is relatively small. Take node 61 as example, as $r/x$ ratio becomes larger, errors of node 54 to 60 declines, also affecting node 1 to 8. It is obvious that LBF can have a lower error with comparable accuracy of power flow when both $r/x$ ratio of branch and $P/Q$ ratio of node load are relatively large. Therefore, LBF model would be suitable for distribution systems with high $P/Q$ ratio in power load/generations, in other words, less reactive power load.

\begin{figure}[ht!]
\captionsetup{singlelinecheck = false, format= hang, justification=raggedright, font=footnotesize, labelsep=space}
\centering
\vspace{-0.46cm}
\setlength{\abovecaptionskip}{-1.5pt}
\setlength{\belowcaptionskip}{-1.5pt}
\includegraphics[width=3.4in]{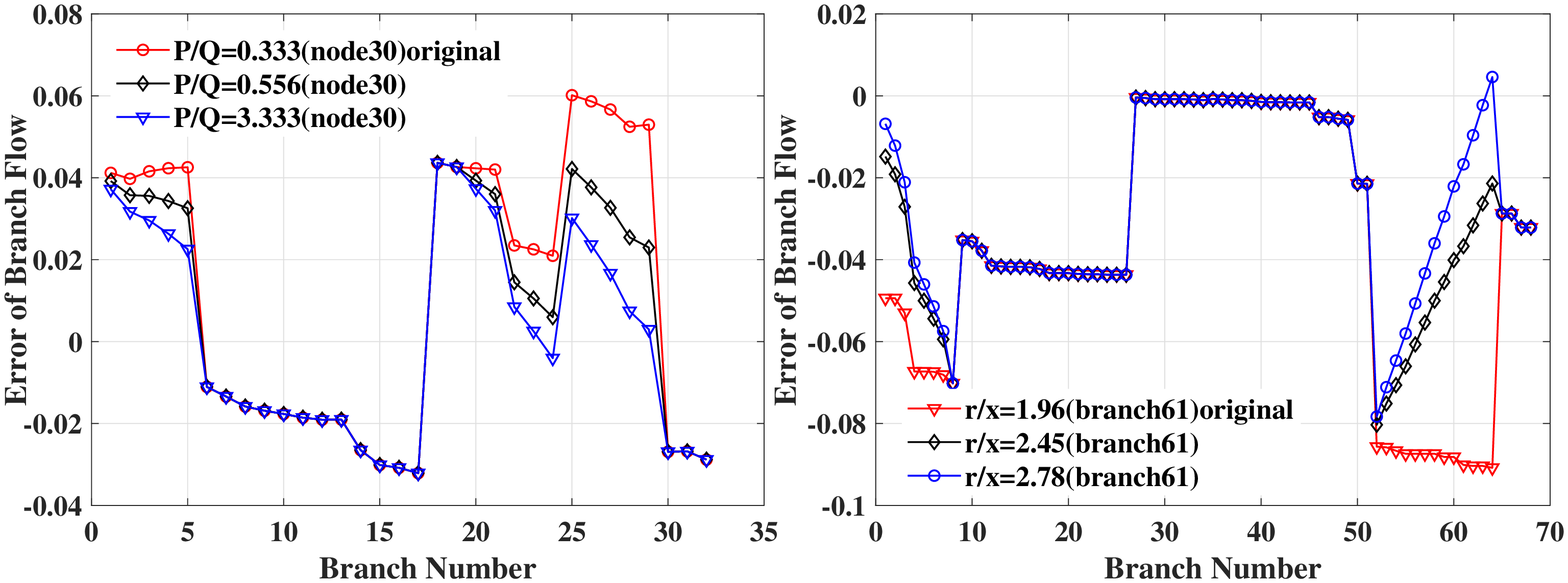}
\caption{Error of Branch Flow Under Different $P/Q$ Ratio Conditions in 33-Bus and $r/x$ Ratio Conditions in 69-Bus Systems}
\label{fig}
\end{figure}\vspace{-7pt}
\begin{figure}[ht!]
\captionsetup{singlelinecheck = false, format= hang, justification=raggedright, font=footnotesize, labelsep=space}
\centering
\vspace{-0.46cm}
\setlength{\abovecaptionskip}{-1.5pt}
\setlength{\belowcaptionskip}{-1.5pt}
\includegraphics[width=3.4in]{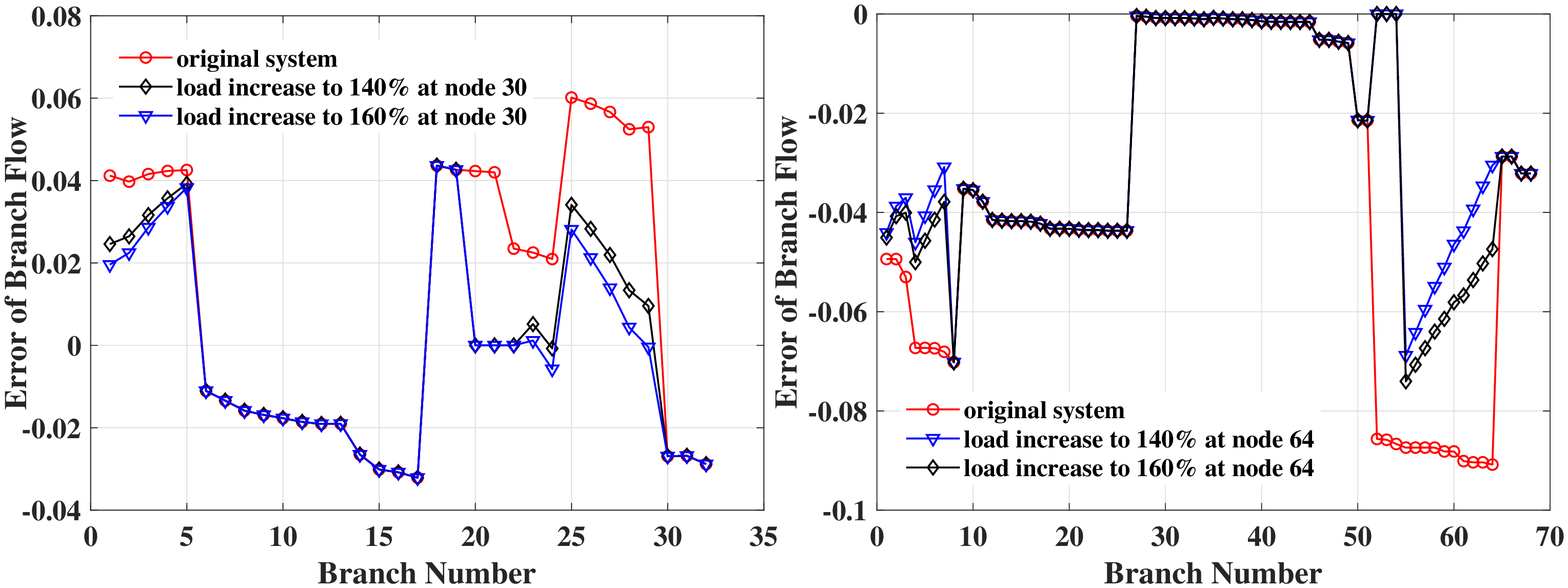}
\caption{Error of Branch Flow Under Different load level Conditions in 33-Bus and 69-Bus Systems}
\label{fig}
\end{figure}\vspace{-5pt}
In Fig.8, we increase power injections at the chosen node with a Static Var Compensator up to 160\%. It is noted that the error declines gradually when the load level increases at the same time. The same situation goes for 69-bus systems. The reason is that when the power injection at the node 30 increases, the reactive power compensation is improved to decrease the reactive currents of branches nearby. Thus, it can make the results of LBF similar to that of ACPF. Therefore, we could conclude that the errors are mainly caused by reactive power flows, which are not precisely calulated in LBF model.

\section{Conclusion}
In this paper, we proposed an in-depth analysis of linearized branch flow (LBF) which is distinguished by its simplicity and high computational efficiency. Besides, we analyzed possible errors and different factors to find the application scope both by theoretical derivation and numerical studies. LBF has a better accuracy in node voltage magnitude compared to branch flow, which is very important in distribution networks. Therefore, LBF is probably suitable for optimization problems of large-scale distribution networks that need high efficiency and high accuracy in voltage magnitude, while the error of branch flows can be accepted in distribution systems where line capacities are large enough, and line losses reduction is not the major concern. Also, given a proper scaling coefficient on current injections of each node, both voltage magnitude and branch flow calculated by LBF could be over-estimations of radial distribution systems. Therefore, LBF can act as a warm start for some complex distribution system optimization problems, e.g. DSEP. Our future research will focus on further applications of LBF in this context.

\bibliographystyle{ieeetr}  
\bibliography{ref} 

\begin{thebibliography}{10}

\bibitem{purchala2005usefulness}
K.~Purchala, L.~Meeus, D.~Van~Dommelen, and R.~Belmans, ``Usefulness of dc
  power flow for active power flow analysis,'' in {\em IEEE Power Engineering
  Society General Meeting, 2005}, pp.~454--459, IEEE, 2005.

\bibitem{sun2012adaptive}
H.~Sun, Q.~Guo, B.~Zhang, W.~Wu, and B.~Wang, ``An adaptive zone-division-based
  automatic voltage control system with applications in china,'' {\em IEEE
  Transactions on Power Systems}, vol.~28, no.~2, pp.~1816--1828, 2012.

\bibitem{yuan2016novel}
H.~Yuan, F.~Li, Y.~Wei, and J.~Zhu, ``Novel linearized power flow and
  linearized opf models for active distribution networks with application in
  distribution lmp,'' {\em IEEE Transactions on Smart Grid}, vol.~9, no.~1,
  pp.~438--448, 2016.

\bibitem{garces2015linear}
A.~Garces, ``A linear three-phase load flow for power distribution system,''
  {\em IEEE Transactions on Power Systems}, vol.~31, no.~1, pp.~827--828, 2015.

\bibitem{gharebaghi2019linear}
S.~Gharebaghi, A.~Safdarian, and M.~Lehtonen, ``A linear model for ac power
  flow analysis in distribution networks,'' {\em IEEE Systems Journal},
  vol.~13, no.~4, pp.~4303--4312, 2019.

\bibitem{haffner2008multistage}
S.~Haffner, L.~F.~A. Pereira, L.~A. Pereira, and L.~S. Barreto, ``Multistage
  model for distribution expansion planning with distributed generation—part
  i: Problem formulation,'' {\em IEEE Transactions on Power Delivery}, vol.~23,
  no.~2, pp.~915--923, 2008.

\bibitem{yang2016state}
J.~Yang, N.~Zhang, C.~Kang, and Q.~Xia, ``A state-independent linear power flow
  model with accurate estimation of voltage magnitude,'' {\em IEEE Transactions
  on Power Systems}, vol.~32, no.~5, pp.~3607--3617, 2016.

\bibitem{teng2003direct}
J.-H. Teng, ``A direct approach for distribution system load flow solutions,''
  {\em IEEE Transactions on power delivery}, vol.~18, no.~3, pp.~882--887,
  2003.

\bibitem{chen2017robust}
X.~Chen, W.~Wu, and B.~Zhang, ``Robust capacity assessment of distributed
  generation in unbalanced distribution networks incorporating anm
  techniques,'' {\em IEEE Transactions on Sustainable Energy}, vol.~9, no.~2,
  pp.~651--663, 2017.

\bibitem{9281770}
T.~{Yang}, Y.~{Guo}, L.~{Deng}, H.~{Shu}, X.~{Shen}, and H.~{Sun}, ``A
  distribution system loss allocation approach based on a modified distflow
  model,'' in {\em 2020 IEEE Power Energy Society General Meeting (PESGM)},
  pp.~1--5, 2020.

\bibitem{baran1989network}
M.~E. Baran and F.~F. Wu, ``Network reconfiguration in distribution systems for
  loss reduction and load balancing,'' {\em IEEE Power Engineering Review},
  vol.~9, no.~4, pp.~101--102, 1989.

\bibitem{wang2017linear}
Y.~Wang, N.~Zhang, H.~Li, J.~Yang, and C.~Kang, ``Linear three-phase power flow
  for unbalanced active distribution networks with pv nodes,'' {\em CSEE
  Journal of Power and Energy Systems}, vol.~3, no.~3, pp.~321--324, 2017.

\bibitem{farivar2013branch}
M.~Farivar and S.~H. Low, ``Branch flow model: Relaxations and
  convexification—part \uppercase\expandafter{\romannumeral1},'' {\em IEEE
  Transactions on Power Systems}, vol.~28, no.~3, pp.~2554--2564, 2013.

\bibitem{gan2014exact}
L.~Gan, N.~Li, U.~Topcu, and S.~H. Low, ``Exact convex relaxation of optimal
  power flow in radial networks,'' {\em IEEE Transactions on Automatic
  Control}, vol.~60, no.~1, pp.~72--87, 2014.

\end{thebibliography}

\end{document}